\documentclass[twocolumn,prd,superscriptaddress,showpacs%
,floatfix,preprintnumbers,nofootinbib,a4paper]{revtex4}

\usepackage{epsfig}

\newcommand{\gag}{g_{a\gamma}}
\long\def\dump#1{}

\newcommand\I{{\rm i}}
\newcommand\E{{\rm e}}

\begin{document}

\title{Signatures of axion-like particles  in
the spectra of TeV gamma-ray sources}

\author{Alessandro Mirizzi}
\affiliation{Max-Planck-Institut f\"ur Physik
(Werner-Heisenberg-Institut),
F\"ohringer Ring 6, 80805 M\"unchen, Germany}
\affiliation{Dipartimento di Fisica and Sezione INFN di Bari,
Via Amendola 173, 70126 Bari, Italy}

\author{Georg G.~Raffelt}
\affiliation{Max-Planck-Institut f\"ur Physik
(Werner-Heisenberg-Institut),
F\"ohringer Ring 6, 80805 M\"unchen, Germany}

\author{Pasquale D.~Serpico}
\affiliation{Center for Particle Astrophysics,
Fermi National Accelerator Laboratory,
Batavia, IL 60510-0500 USA}

\date{23 April 2007}

\preprint{FERMILAB-PUB-07-082-A, MPP-2007-44}

\begin{abstract}
One interpretation of the unexplained signature observed in the
PVLAS experiment invokes a new axion-like particle (ALP) with a
two-photon vertex, allowing for photon-ALP oscillations in the
presence of magnetic fields. In the range of masses and couplings
suggested by PVLAS, the same effect would lead to a peculiar dimming
of high-energy photon sources. For typical parameters of the
turbulent magnetic field in the galaxy, the effect sets in at
$E_\gamma\agt10~{\rm TeV}$, providing an ALP signature in the
spectra of TeV gamma sources that can be probed with Cherenkov
telescopes. A dedicated search will be strongly motivated if the
ongoing photon regeneration experiments confirm the PVLAS particle
interpretation.
\end{abstract}

\maketitle

\section{Introduction}
\label{Introduction}

One of the phenomenologically most important properties of the
hypothetical axions is their two-photon vertex that allows for
axion-photon conversions in external electric or magnetic
fields~\cite{Dicus:1978fp, sikivie}. In particular, this coupling is
used by the ADMX experiment to search for axion dark
matter~\cite{Bradley:2003kg, Duffy:2006aa} and by the CAST
experiment to search for solar axions~\cite{Zioutas:2004hi,
Andriamonje:2007ew}.  Generically such particles affect the
propagation of photons in magnetic fields. For a linearly polarized
laser beam propagating in a transverse $B$-field, signatures are a
rotation of the plane of polarization and the development of an
elliptical polarization component~\cite{Maiani:1986md,
Gasperini:1987da, Raffelt:1987im}. The latter effect is also caused
by the effective four-photon interaction predicted by
QED~\cite{adler, Iacopini:1979ci}.

Recently the laser experiment PVLAS has reported such results with
an amplitude about $10^4$ times larger than expected from
QED~\cite{Zavattini:2005tm}. If one interprets this signal in terms
of photon-axion conversions, these measurements imply an axion mass
$m_a\approx 1.3$~meV and a coupling with photons
$g_{a\gamma}\approx3\times 10^{-6}~{\rm GeV}^{-1}$, where the
coupling constant is defined in Eq.~(\ref{pscoupl}) below. This
combination of $m_a$ and $\gag$ is incompatible with axions in the
usual sense. Therefore, the new states require a different
interpretation and are generically termed ``axion-like particles''
(ALPs), meaning bosons with a two-photon vertex where the mass and
coupling strength are taken as independent parameters.

The main problem with the PVLAS signature is that it violates simple
astrophysical limits by a huge margin. ALPs are produced in the Sun
and other stars by the Primakoff process where thermal photons
convert in the fluctuating electric fields of the stellar
plasma~\cite{Dicus:1978fp, Raffelt:1985nk, Raffelt:1999tx}. Assuming
the PVLAS-inspired parameters, a standard solar model leads to an
ALP luminosity so large that the Sun would burn out in 1000~years.
Circumventing this vast discrepancy is the main theoretical
challenge for the PVLAS particle interpretation~\cite{Masso:2005ym,
  Masso:2006gc, Mohapatra:2006pv, Coriano:2006xh, Fukuyama:2006hj,
  Antoniadis:2006wp}.

It is conceivable that the presence of the hot stellar plasma
modifies the effective couplings or that these couplings are
different at the momentum transfers relevant in stars. Therefore, it
has been stressed that the PVLAS particle interpretation should be
tested with experiments where the transition takes place in vacuum
and where the momentum transfer is small~\cite{Jaeckel:2006xm}.
Photon regeneration experiments (``shining light through a wall'')
are of particular interest because it will be fairly easy to confirm
PVLAS if the particle interpretation is indeed correct. Several such
efforts are now being discussed or are already under
way~\cite{Ringwald:2006rf, princeton, cerncourier}, notably ALPs at
DESY,
 BMV at LULI, GammeV at Fermilab, LIPSS at
Jefferson Laboratory, OSQAR at CERN, and PVLAS-regeneration at INFN
Laboratory in Legnaro.

If the PVLAS particle interpretation is confirmed, some radical new
low-energy physics must be at work that prevents ALP emission from
stars. However, other astrophysical settings provide conditions
similar to the laboratory experiments, i.e., a vacuum environment
and near-vanishing momentum transfers. One example is ``shining
light through the Sun'' where a high-energy photon source would
become visible through the Sun by photon-ALP conversion in the solar
magnetic field on the far side of the Sun, and their regeneration on
our side~\cite{Fairbairn:2006hv}. Another example is the double
pulsar J0737-3039, where gamma rays emitted by one pulsar
periodically pass through the magnetosphere of the other on their
way to us~\cite{Dupays:2005xs}.

We here consider another example, the photon-ALP conversion in the
turbulent magnetic field of our galaxy. Beyond energies of order
10~TeV, the gamma-ray flux would be depleted, leaving a distinct
signature in the spectrum of TeV gamma-ray sources. Current data
from Imaging Atmospheric Cherenkov Telescopes (IACTs) do not allow
for a stringent constraint on this effect. However, if the
laboratory experiments confirm the existence of ALPs with the
properties suggested by PVLAS, this depletion must be included in
the analysis of TeV gamma-ray sources by IACTs.  Given the strong
motivation that would be provided by a positive laboratory ALP
confirmation, dedicated efforts by present and future instruments
would be mandatory that could provide an independent astrophysical
signature of these novel particles and/or allow one to study or
constrain the turbulent galactic $B$~field.

We begin in Sec.~\ref{Formalism} with a summary of the formalism to
describe photon-ALP conversions and turn in Sec.~\ref{randomB} to
phenomenological consequences on the propagation of TeV photons in
our galaxy. In Sec.~\ref{MCP} we briefly touch on the possible
effect of millicharged particles on photon propagation. We conclude
in Sec.~\ref{conclusion}.

\section{Photon-axion conversion}
\label{Formalism}

Axion-like particles by definition have a two-photon coupling. For
pseudoscalars, it is of the form
\begin{equation}\label{pscoupl}
 {\cal L}_{a\gamma} =
 -\frac{1}{4} \gag F_{\mu\nu}\tilde{F}^{\mu\nu}a
 =\gag{\bf E}\cdot{\bf B}\,a\,,
\end{equation}
where $a$ is the axion-like field with mass $m_a$, $F_{\mu\nu}$ the
electromagnetic field-strength tensor, $\tilde{F}_{\mu\nu}
\equiv\frac{1}{2}\epsilon_{\mu\nu\rho\sigma}F^{\rho\sigma}$ its
dual, and $\gag$ the ALP-photon coupling with dimension of inverse
energy. For a scalar particle, the coupling is proportional to
$F_{\mu\nu}F^{\mu\nu}a$. To be definite we limit our discussion to
the pseudoscalar case, but similar consequences apply to scalars.

As a consequence of this coupling, ALPs and photons oscillate into
each other in an external magnetic field. Under quite general
assumptions, the probability for an unpolarized photon beam to
convert to ALPs after traversing a magnetic field ${\bf B}
=(B_x,B_y,B_z)$ from $0$ to $z$ is (Appendix~A)
\begin{eqnarray}\label{eq:conversion1}
 P_{\gamma \to a}(z)&=&
 \frac{\gag^2}{8}\bigg(\left|\int_0^z{\rm d}z^{\prime}\,
 {\rm e}^{-{\I}2\pi z^{\prime}/l_0}B_{x}(x,y,z^{\prime})\right|^2
 \nonumber\\
 &&{}+
 \left|\int_0^z{\rm d}z^{\prime}\, {\rm e}^{-{\I}2\pi z^{\prime}/l_0}
 B_{y}(x,y,z^{\prime})\right|^2\bigg)\,,
 \label{equatnr1}
\end{eqnarray}
where for simplicity we have chosen the $z$--axis along the
propagation direction. Further, $l_0=4\pi E/m_a^2$ is the
oscillation length with $m_a$ the axion mass and $E$ the photon
energy. The  meV range of ALP masses, relevant for the PVLAS
particle interpretation,  is so large that the photon plasma mass is
completely negligible by comparison, in contrast to the case of
cosmic microwave conversion into intergalactic magnetic fields,
studied in~\cite{Mirizzi:2005ng, Mirizzi:2006zy}.

We consider a simplified case where the field is of constant
magnitude and random direction in each patchy domain, each with
typical size $s\ll z$, so that a large number $N$ of domains is
crossed. The previous expression then further simplifies to
(Appendix~\ref{derivation})
\begin{equation}\label{pertpz1}
 P_{\gamma \to a}(z)=N\,P_0\,,
\end{equation}
where the probability per single domain is
\begin{equation}\label{pertpz1a}
 P_0\approx \frac{\gag^2\langle|{\bf B}|^2\rangle\,s^2}{4}\,
 \frac{\sin^2(\pi\,s/l_0)}{(\pi\,s/l_0)^2}\,.
\end{equation}
Equation~(\ref{pertpz1}) only holds in the perturbative regime where
$N\,P_0\ll 1$. For $N$ sufficiently large, this result violates
unitarity. It can be shown (Appendix of Ref.~\cite{Mirizzi:2006zy})
that the correct continuum limit after travelling over $z\gg s$ is
\begin{equation}\label{totprob}
 P_{\gamma \to a}(z) = {1 \over 3}
 \left[1-\exp\left(-\frac{3P_0\,z}{2s}\right)\right]\,.
\end{equation}
As physically expected, Eq.~(\ref{totprob}) implies for
$z/s\to\infty$ that the conversion probability saturates so that on
average one third of all photons converts to axions.

\section{Conversions in the turbulent galactic magnetic field}
\label{randomB}

For the PVLAS-inspired parameters $m_a=1.3$~meV and
$\gag=3\times10^{-6}~{\rm GeV}^{-1}$, it is useful to write $P_0$ in
suitable numerical units,
\begin{eqnarray} \label{estim}
 P_0 &=& (1.5\,g_{6}\,B_{\rm \mu G}\,s_{\rm pc})^2\,
 \frac{\sin^2(3.8{\times}10^{3}\, m_{\rm meV}^2\,s_{\rm pc}
 /E_{10})}{(3.8{\times}10^{3}\, m_{\rm meV}^2\,s_{\rm pc}/E_{10})^2}
 \nonumber \\
 &\simeq& 0.8\times10^{-7}
 \left(\frac{g_{6}\,B_{\rm \mu G}\,E_{10}}{m_{\rm meV}^2}\right)^2\,.
\end{eqnarray}
Here, we have introduced $g_{6}=\gag/10^{-6}~{\rm GeV}^{-1}$,
$B_{\rm\mu G}$ is the root mean square (rms) magnetic field strength
in micro-Gauss, $s_{\rm pc}$ is the domain size in pc, $m_{\rm meV}$
is the ALP mass in meV, and $E_{10}$ the photon energy in units of
10~TeV. In the second line we have replaced $\sin^2$ with its
average value $\frac{1}{2}$ because its argument is large and
oscillates rapidly for any realistic energy resolution.

Although the galactic $B$ field has a regular component with several
kpc coherence length, on small scales a turbulent component
dominates (Ref.~\cite{Han:2004aa} and references therein). The power
spectrum follows a Kolmogorov power law, with a lower cutoff at very
small dissipative scales, perhaps as small as $6\times
10^{-4}$~pc~\cite{McIvor}, but in any case at most comparable to
0.01~pc, with an rms intensity of order $\mu$G on pc scales.  For
typical galactic distances of 10~kpc, there are approximately $10^6$
domains with $s \approx 0.01$~pc towards a typical TeV gamma source
such as the one at the galactic center~\cite{Tsuchiya:2004wv,
Kosack:2004ri, Aha, Albert:2005kh}. For nominal values of the
parameters in Eq.~(\ref{estim}), $P_0\simeq 10^{-7}$--$10^{-6}$ at
energies of order 10~TeV, implying $N\,P_0\simeq0.1$--1. Therefore,
observable effects must be expected.

\begin{figure}
\centering \epsfig{figure=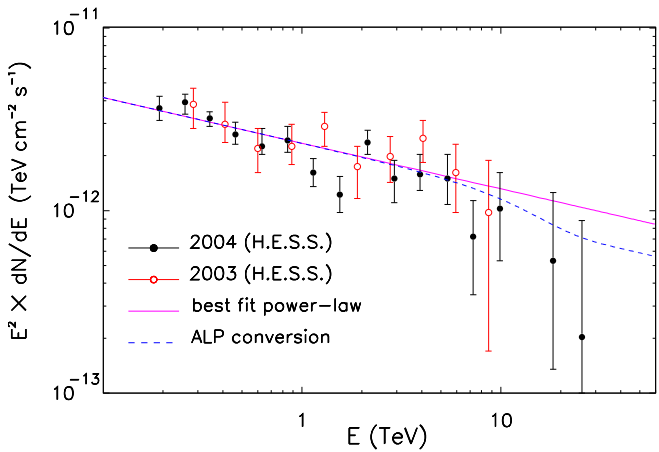,width=1.\columnwidth,angle=0}
\caption{\label{fig1} Spectral energy density $E^2\times
\mathrm{d}N/\mathrm{d}E$ of photons from the galactic center source,
for the 2004 data (full points) and 2003 data (open points) of
H.E.S.S.~\cite{Aha2006}.  Error bars represent 95\% CL.  The
continuous line shows the best-fit power-law
$\mathrm{d}N/\mathrm{d}E \sim E^{-\Gamma}$ with $\Gamma =
2.25$~\cite{Aha2006}. The dashed line shows the effect of photon-ALP
conversion with  coupling and mass suggested by PVLAS.}
\end{figure}

In Fig.~\ref{fig1} we show the spectral modification of a TeV source
at the galactic center (distance 8.5~kpc) superimposed with
H.E.S.S.~data~\cite{Aha2006}. For illustration we have used
Eq.~(\ref{totprob}) with $g_{6}=3$, $B_{\rm \mu G}=0.7$, $s_{\rm
pc}=0.01$ and $m_{\rm meV}=1$ and we have assumed that the power-law
spectrum does not break before 60 TeV. Photon-ALP oscillations
(dashed curve) cause a downward shift of the spectrum at high
energies, i.e., a change of normalization of the typical power-law
spectrum (continuous curve) between low and high energies
($E\agt10$~TeV). The maximum shift is 33\% when the conversion
saturates.

Evidently current data do not allow for a serious constraint on this
depletion effect. Note, however, that the large error bars at high
energy are only due to a lack of statistics. The points reported in
Fig.~\ref{fig1} are based on 17 hours of data in 2003 with two
telescopes and 48.7 hours in 2004 in the four-telescope array mode.
Already the current generation of IACTs (\hbox{CANGAROO-III},
H.E.S.S., MAGIC, VERITAS) may have a sufficient aperture to probe
this scenario, if dedicated campaigns were motivated by a positive
laboratory detection.

To be more quantitative, one would model the turbulent field as a
Gaussian random field with zero mean and an rms value $B_{\rm
rms}$~\cite{Han:2004aa}.  Each of its components can thus be written
in terms of its Fourier transform as
\begin{equation}
 B_i({\bf x})=\int\frac{{\rm d}^3 {\bf k}}{(2\pi)^3}\,
 \tilde{B}_i({\bf k})\,
 {\rm e}^{{\I}[{\bf x}\cdot{\bf k}+\phi_i({\bf k})]}
\end{equation}
where the phases $\phi_i({\bf k})$ are random.  For an isotropic and
homogeneous turbulence, the Fourier modes satisfy
\begin{equation}
 \langle\tilde{B}_i({\bf k})\tilde{B}^{*}_j({\bf k}^{\prime})\rangle
 =\frac{{\cal B}^2(k)}{8\pi k^2}\,
 \left(\delta_{ij}-\frac{k_i\,k_j}{k^2}\right)
 (2\pi)^6\delta^3({\bf k}-{\bf k}^{\prime}),
\end{equation}
where the tensor in brackets implements the condition
$\nabla~\cdot~{\bf B}=0$. In the generic case of a power-law
spectrum with index $\alpha$ between the scales $s_{\rm min}$ and
$s_{\rm max}$, i.e., between wavenumbers $k_{\rm L}\equiv
2\pi/s_{\rm max}$ and $k_{\rm H}\equiv 2\pi/s_{\rm min}$, one has
\begin{equation}
 {\cal B}^2(k)=B_{\rm rms}^2\,(\alpha-1)\,k^{-\alpha}
 \left(k_{\rm L}^{1-\alpha}-k_{\rm H}^{1-\alpha}\right)^{-1}\,,
\end{equation}
which is already normalized such that $\langle|{\bf B}({\bf
x})|^2\rangle=B_{\rm rms}^2$. In the limit $k_{\rm L}\ll k_{\rm H}$,
and if $\alpha>1$, one finds
\begin{equation}
 {\cal B}^2(k)\simeq B_{\rm rms}^2\,(\alpha-1)\,
 k^{-\alpha} k_{\rm L}^{\alpha-1}\,.
\end{equation}
Therefore, the field averaged over scales less than $s$ is
\begin{equation}
 \langle |{\bf B}({\bf x})|^2\rangle_s=
 B_{\rm rms}^2\,(s/s_{\rm max})^{\alpha-1}\,.
\end{equation}
For the Kolmogorov spectrum $\alpha=5/3$ suggested by the data, this
means that the rms intensity of the field varies as $s^{1/3}$.  The
intensity below 0.001~pc is then only a factor 10 weaker than the
$\mu$G level at the pc scale. Below 0.01~pc it is only a factor
$\sim4$ lower than at the pc scale. Therefore, our simple estimate
of $P_0$ may be too optimistic by an order of magnitude. However,
the effect would still be observable simply by looking at a factor
$\sim 3$ larger energies. Additionally, the true field configuration
may be more complicated, and recently a more intense turbulence than
previously estimated has been suggested \cite{Hantalk}.

In a more detailed treatment one would consider stochastic
realizations of the realistic power spectrum of the turbulent $B$
field. However, for our purpose simple estimates are probably more
instructive and show that: (i)~Possible effects may start
manifesting themselves around 10~TeV, and are more and more likely
to show up at 20--30~TeV. (ii)~The smaller the characteristic scale
of turbulence of a given intensity, the larger the number of domains
available, and the lower the energy at which the effect appears.
(iii)~The conversion probability depends on~$E^2$. Therefore, on the
scale of the typical energy resolution of a Cherenkov telescope, the
depletion rapidly drops from negligible to the saturation value
of~${1}/{3}$. (iv)~The phenomenology described here would be
universal, affecting both galactic and extragalactic sources. Yet,
the exact energy at which the shift manifests depends on the
properties of the field along that line of sight. Although for all
sources the light must cross the galactic B field to reach us, one
may not exclude an additional role of a small-scale field close to
the sources. Our estimate for the onset of the effect is
conservative, especially for extragalactic sources. As a general
rule, for sources in similar directions, the more distant ones may
manifest the signature at lower energies.

\section{Millicharged particles}
\label{MCP}

Another particle-physics explanation of the PVLAS anomaly postulates
the existence of low-mass milli\-charged
particles~\cite{Gies:2006ca,Ahlers:2006iz}. We briefly check if this
hypothesis would also affect the propagation of photons in the
astrophysical context.

At TeV energies, the extragalactic medium becomes opaque due to the
onset of $e^{\pm}$ pair-production on the diffuse low-energy photon
backgrounds. At a few PeV, the mean free path of photons reaches a
minimum of $\lambda_e \lesssim 10$~kpc due to pair production on the
Cosmic Microwave Background (CMB)~\cite{Lee:1996fp}. The threshold
energy $E_{\rm th}\sim 3 \times 10^{14}$~eV scales as $m_e^2$ and
the cross section as $e^4/m_e^2$. Scaling these quantities to
millicharged particles with charge $q\ll e$ and mass $m_q\ll m_e$
one finds
\begin{equation}
 \lambda_q^{-1}\simeq \lambda_e^{-1}
 \left(\frac{q}{e}\right)^4\left(\frac{m_e}{m_q}\right)^2,
\end{equation}
and
\begin{equation}
 E_{\rm th}^q\simeq E_{\rm th}^e\left(\frac{m_q}{m_e}\right)^2\,.
\end{equation}
The preferred mass range of the millicharged candidate is
0.01--0.1~eV, i.e., $m_q/m_e\sim 2\times 10^{-8}$--$2 \times
10^{-7}$. The peak of the cross section is very close to the
threshold and would fall in the $(10^{-16}$--$10^{-14})\times
10^{15}$~eV range, i.e., ranging from infrared to ultraviolet.
Sources at cosmological distances do not show such a universal
dimming. The conservative requirement $\lambda_q\agt 1$~Gpc implies
$q\alt 10^{-5}\,e$.

A much more constraining limit of $q \alt 10^{-7} e$ arises from
spectral distortion effects of the CMB that may already rule out the
millicharged particle explanation of PVLAS~\cite{Melchiorri:2007}.
In any event, it appears safe to assume that millicharged particles
with the relevant properties would not affect TeV photon
observations.

\section{Conclusions}
\label{conclusion}

The unexpected optical properties of the vacuum suggested by the
PVLAS experiment has inspired various interpretations in terms of
axion-like particles. The severe conflict with stellar structure
arguments implies that this interpretation requires radical new
physics at low energies. If the new particles interact differently
in a stellar plasma or at vanishing momentum transfers, they may
still show up in the upcoming photon regeneration experiments. In
this case one necessarily expects signatures also in other settings
that are characterized by a vacuum environment and/or small momentum
transfers.

We have discussed possible signatures of PVLAS particles in the
spectra of TeV gamma-ray sources in our galaxy.  If the PVLAS signal
can be attributed to photon-ALP conversion in the laboratory, the
same effect must occur in the astrophysical context. For an ALP mass
around 1~meV, as suggested by PVLAS, one would observe a peculiar
distortion in the photon spectra at $E_\gamma\agt10~{\rm TeV}$ due
to conversions in the turbulent galactic $B$-field. This process
would take place under better vacuum conditions than are achievable
in the laboratory and the momentum transfer would be extremely
small.

Present data from TeV gamma-ray telescopes do not allow for a
stringent constraint on this effect. However, a positive ALP
detection would strongly motivate a dedicated search, perhaps
allowing one to find signatures for ALPs in current or future
instruments and to investigate or constrain the properties of the
turbulent magnetic field in the galaxy and beyond.

\section*{Acknowledgments}

We would like to thank Marco Roncadelli and Pratik Majumdar for
comments. P.S.~acknowledges support by the US Department of Energy
and by NASA grant NAG5-10842. The work of A.M. is supported by an
Alexander von Humboldt fellowship grant. A.M.~also acknowledges
support in the initial phase of this work by the Italian ``Istituto
Nazionale di Fisica Nucleare'' (INFN) and by the ``Ministero
dell'Istruzione, Universit\`a e Ricerca'' (MIUR) through the
``Astroparticle Physics'' research project. In Munich, partial
support by the Deutsche Forschungsgemeinschaft Grant TR~27 by the
Cluster of Excellence ``Origin and Structure of the Universe''
(Garching and Munich), and by the European Union under the ILIAS
project, contract No.~RII3-CT-2004-506222, is acknowledged.

\appendix
\section{Photon-ALP conversion in a random
  magnetic field}\label{derivation}

We here derive the photon-ALP conversion probabilities
[Eqs.~(\ref{eq:conversion1})--(\ref{pertpz1})] in a random magnetic
field distribution. These detailed results are not used for the
simple estimates derived in our paper, but would be necessary for a
detailed treatment involving the numerical study of different
realizations of the turbulent galactic B-field.

For relativistic ALPs, the equations of motion following from
Eq.~(\ref{pscoupl}) reduce to the linearized
system~\cite{Raffelt:1987im}
\begin{equation}\label{linsys1}
\left(\omega-{\I}\partial_z +{\cal M}\right) \left(
\begin{array}{ccc}
A_x\\
A_y\\
a
\end{array}\right)=0,
\end{equation}
where $z$ is the direction of propagation, $A_x$ and $A_y$ are
orthogonal components of the photon field in a fixed frame
perpendicular to $z$, and $\omega$ is the photon energy. The mixing
matrix~is
\begin{equation}  \label{mixmatxy}
{\cal M}=\left(
  \begin{array}{ccc}
    \Delta_{xx}&\Delta_{xy}& \frac{1}{2}\gag B_x\\
    \Delta_{yx}&\Delta_{yy}&\frac{1}{2}\gag B_y\\
    \frac{1}{2}\gag B_x&\frac{1}{2}\gag B_y& \Delta_a
  \end{array}
  \right),
\end{equation}
where $\Delta_a = -m^2_{a}/2\omega$. Notice that the component of
${\bf B}$ parallel to the direction of motion does not induce
photon-axion mixing, since only $B_x$ and $B_y$ enter the third
row/column of ${\cal M}$. The entries $\Delta_{ij}$ ($i,j=x,y$) that
mix the photon polarization states are energy-dependent terms
determined by the properties of the medium and the QED vacuum
polarization effect. We will neglect the latter because it is
sub-dominant here.

The $\Delta_{ij}$ terms have a simple interpretation when the $x$ or
$y$ direction coincides with the transverse field direction ${\bf
B}_{\rm T}= {\bf B}-({\bf B}\cdot \hat{{\bf z}})\,\hat{{\bf z}}$. We
can then specify the previous equations for the case of a single
domain with uniform magnetic field ${\bf B}_{\rm T}$, whose
modulus  will be denoted by $B_T=|\bf{B}_T|$. Equation~(\ref{linsys1}) is in the
new basis
\begin{equation}  \label{linsys}
\left(\omega-{\I}\partial_z +{\cal M}\right) \left(
\begin{array}{ccc}
A_{\perp}\\
A_{\parallel}\\
a
\end{array}
\right)=0,
\end{equation}
where $i={\perp}$ or ${\parallel}$ refer to the ${\bf B}_{\rm T}$
direction. The mixing matrix is now
\begin{equation}  \label{mixmat}
{\cal M}\equiv\left(
\begin{array}{ccc}
\Delta_{\perp} &\Delta_{\rm R}     & 0\\
\Delta_R       &\Delta_{\parallel} & \Delta_{a\gamma}\\
0              &\Delta_{a\gamma}   & \Delta_a\\
  \end{array}
  \right),
\end{equation}
where
\begin{equation}\label{eq12}
\begin{array}{cccccc}
\Delta_{\perp}&=&\Delta_{\rm pl}+\Delta_{\perp}^{\rm CM}\,,& \quad
\Delta_{a \gamma}&=& \frac{1}{2}\gag B_{\rm T}\,,\\
\Delta_{\parallel}&=&\Delta_{\rm pl}+\Delta_{\parallel}^{\rm CM}\,,
&\quad \Delta_{\rm pl}&=& -\omega_{\rm pl}^2/2\omega\,.
\end{array}
\end{equation}
Here, $\omega_{\rm pl}^2 = 4\pi\alpha\,n_e/m_e$ is the plasma
frequency with $m_e$ the electron mass and $\alpha$ the
fine-structure constant. The Faraday rotation term $\Delta_{\rm R}$,
which depends on the energy and the longitudinal component $B_z$,
would couple the modes $A_{\parallel}$ and $A_{\perp}$. While it is
important when analyzing polarized photon sources, it plays a
negligible role here. The $\Delta_{\rm CM}$ terms describe the
Cotton-Mouton effect, i.e., the birefringence of fluids in presence
of a longitudinal magnetic field, with $|\Delta_{\parallel}^{\rm
CM}-\Delta_{\perp}^{\rm CM}|\propto B_{\rm T}^2$. These terms are of
little importance for the following arguments and will be neglected
hereafter.

Therefore, we finally concentrate on the simple two-level mixing
problem
\begin{equation}
 \left[\omega-{\I}\partial_z +
 \left(
  \begin{array}{cc}
    \Delta_{\rm pl}&\Delta_{a \gamma}\\
    \Delta_{a \gamma}&\Delta_a
  \end{array}
 \right)\right]
 \left(
 \begin{array}{cc}
 A_\parallel\\a
 \end{array}
 \right)=0\,.
\end{equation}
The solution of this system follows from a diagonalization of the
mixing matrix by a rotation with an angle
\begin{equation}\label{tan}
 \vartheta = \frac{1}{2}\arctan\left(\frac{2\Delta_{a
 \gamma}}{\Delta_{\rm pl}-\Delta_a}\right).
\end{equation}
In analogy to the neutrino case, the probability for a photon
emitted in the state $A_{\parallel}$ to convert to an ALP after
traveling a distance $s$ in a constant transverse magnetic field
${\bf B}_{\rm T}$ is
\begin{eqnarray}
  P_0(\gamma\rightarrow a)&=& \left|\langle A_\parallel(0)\mid a(s)
  \rangle\right|^2 \label{p1ga00} \\
  &=&\sin^2\left(2 \vartheta \right)\sin^2\left(
  \Delta_{\rm osc}s/2\right)  \label{p1ga0}\\
  &=&\left(\Delta_{a \gamma} s\right)^2 {\sin^2(\Delta_{\rm osc} s /2)
  \over (\Delta_{\rm osc} s /2)^2} \;\ ,  \label{p1ga}
\end{eqnarray}
where the oscillation wave number is given by
\begin{equation}\label{deltaosc}
 \Delta_{\rm osc}^2=
 (\Delta_{\rm pl}-\Delta_a)^2 + 4 \Delta_{a\gamma}^2 \,.
\end{equation}
The conversion probability is energy independent when
$2|\Delta_{a\gamma}|\gg|\Delta_{\rm pl}-\Delta_{a}|$ or in any case
when the oscillatory term $\sin^2x/x^2\approx 1$ in
Eq.~(\ref{p1ga}), corresponding to $\Delta_{\rm osc} s /2~\ll~1 $.

We now return to the $3{\times}3$ formalism to derive a perturbative
solution. In a fixed $x$-$y$-$z$ frame with $z$ the direction of
motion, the propagation equations are
\begin{equation}\label{fulleq}
\left[\omega-{\I}\partial_z + \left(
  \begin{array}{ccc}
    \Delta_{xx}&\Delta_{xy}& \Delta_{a\gamma}\,s_{\gamma}\\
    \Delta_{yx}&\Delta_{yy}&\Delta_{a\gamma}\,c_{\gamma}\\
    \Delta_{a\gamma}\,s_{\gamma}&\Delta_{a\gamma}\,c_{\gamma}& \Delta_a
  \end{array}
  \right)
\right] \left(
\begin{array}{ccc}
A_x\\
A_y\\
a
\end{array}
\right)=0\,,
\end{equation}
where $c_\gamma=\cos\gamma$ and $s_\gamma=\sin\gamma$ with $\gamma$
the angle between ${\bf B}_T$ and the $y$ axes (measured clockwise).
Further, from Eq.~(A5) one can write
\begin{eqnarray}
    \Delta_{xx}&\simeq &\Delta_{\rm pl} \,\ ,
    \nonumber\\
    \Delta_{xy}&\simeq & \,\  0   \,\  \,\ ,
    \nonumber\\
    \Delta_{yy}&\simeq &\Delta_{\rm pl} \,\ .
\end{eqnarray}
The field strength entering $\Delta_{a\gamma}$ is $B_{\rm T}=|{\bf
B}_{\rm T}|=|{\bf B}|\,|\sin\psi|$, where $\psi$ is the angle
between the field and the photon propagation direction. Thus we have
$B_x=B_{\rm T}c_\gamma$, $B_y=B_{\rm T}s_\gamma$ that are all
$z$-dependent quantities. All of the $\Delta_{ij}$ are $z$-dependent
as well because this applies to $\gamma$, $n_e$, and $B_T$,
entering the quantities in Eq.~(\ref{eq12}).

Since the ALP is weakly coupled, the 3rd row/column off-diagonal
terms are much smaller than $\omega$, and it makes sense to write
\begin{equation}
{\I}\partial_z{\bf A}=({\mathcal H}_0+{\mathcal H}_1){\bf A}
\end{equation}
where ${\bf A}=(A_x,A_y,a)$,
\begin{equation}
{\cal H}_0 =\omega\;{\bf I}+\left(
  \begin{array}{ccc}
    \Delta_{\rm pl}& 0 & 0\\
    0 &\Delta_{\rm pl}&0\\
     0         &      0    & \Delta_a
  \end{array}
  \right)\,,
\end{equation}
and
\begin{equation}
{\cal H}_1=\left(
  \begin{array}{ccc}
    0 & 0 & \Delta_{a\gamma}\;s_{\gamma}\\
    0 & 0 & \Delta_{a\gamma}\;c_{\gamma}\\
    \Delta_{a\gamma}\;s_{\gamma} & \Delta_{a\gamma}\;c_{\gamma} & 0
  \end{array}
  \right).
\end{equation}
For $\gag\to 0$ this equation is solved exactly by ${\bf
A}^{(0)}(z)={\mathcal U}_0(z){\bf A}(0)$, where
\begin{equation}
 {\mathcal U}_0(z)=
 \exp\left[-{\I}\int_{0}^z{\rm d}z^{\prime}
 {\mathcal H}_0(z^{\prime})\right]\,.
\end{equation}

If we now include the perturbation, the complete solution can be
written perturbatively in the interaction representation. In
particular, to first order we have ${\bf A}_{\rm int}={\mathcal
U}_0^{\dagger}{\bf A}$, ${\mathcal H}_{\rm int}={\mathcal
U}_0^{\dagger}{\mathcal H}_1{\mathcal U}_0$, and
\begin{equation}
 {\bf A}_{\rm int}^{(1)}(z)=-{\I}
 \int_0^z{\rm d}z^{\prime}\,
 {\mathcal H}_{\rm int}(z^{\prime}){\bf A}_{\rm int}^{(0)}(0),
\end{equation}
and ${\bf A}_{\rm int}^{(0)}(z)={\bf A}(0)$ because
\begin{equation}
 {\bf A}_{\rm int}^{(0)}(z)=
 {\mathcal U}_0^{\dagger}{\bf A}^{(0)}(z)=
 {\mathcal U}_0^{\dagger}{\mathcal U}_0{\bf A}(0)\,.
\end{equation}
Since
${\mathcal H}_0$ is diagonal,
 ${\mathcal U}_0$ has the general form
${\mathcal U}_0(z)={\rm
diag}[\E^{-{\I}a(z)},\E^{-{\I}b(z)},\E^{-{\I}c(z)}]$ so that

\begin{equation}
{\mathcal H}_{\rm int}= \left(
  \begin{array}{ccc}
    0 & 0 & \E^{-{\I}(c-a)}\Delta_{a\gamma}s_{\gamma}\nonumber\\
    0 & 0 & \E^{-{\I}(c-b)} \Delta_{a\gamma}c_{\gamma}\nonumber\\
   \E^{{\I}(c-a)} \Delta_{a\gamma}s_{\gamma} &
   \E^{{\I}(c-b)}\Delta_{a\gamma}c_{\gamma} & 0\nonumber
  \end{array}
  \right).
\end{equation}
The ALP amplitude developed at distance $z$ is then
\begin{widetext}
\begin{equation}
 a^{(1)}(z)= -{\I}\,\frac{\gag}{2}
 \int_0^z{\rm d}z^{\prime}
 \left\{A_x(0)\,B_x(z^{\prime})\E^{{\I}[c(z^{\prime})-a(z^{\prime})]}
 +A_y(0)\,B_y(z^{\prime})\E^{{\I}[c(z^{\prime})-b(z^{\prime})]}
 \right\}\,.
\end{equation}
This result is a straightforward generalization of the one derived
in Ref.~\cite{Raffelt:1987im}. The probability for photon-ALP
conversion is then schematically
\begin{equation}
 P_{\gamma \to a}(z)=|A_x(0)|^2 |{\mathcal I}_1|^2
 +|A_y(0)|^2\,|{\mathcal I}_2|^2
 +2\,{\rm Re}[A_x(0)A_y(0)\,{\mathcal I}_1\, {\mathcal I}_2] .
\end{equation}
For an unpolarized source, an average over the initial state has to
be performed. The interference term averages to zero, and
$\langle|A_x(0)|^2\rangle =\langle|A_y(0)|^2 \rangle=1/2$. Then
\begin{equation}\label{BxBy}
 P_{\gamma \to a}(z)=\frac{\gag^2}{8}
 \left(\left|\int_0^z{\rm d}z^{\prime}
 \E^{{\I}(\Delta_a-\Delta_{\rm pl})\,z^{\prime}}B_{x}(z^{\prime})
 \right|^2+
 \left|\int_0^z{\rm d}z^{\prime}
 \E^{{\I}(\Delta_a-\Delta_{\rm pl})\,
 z^{\prime}}B_{y}(z^{\prime})\right|^2\right)
\end{equation}
or equivalently
\begin{equation}\label{eq:prob2}
 P_{\gamma \to a}(z) =
 \frac{\gag^2 |{\bf B}|^2}{8}
 \left(\left|\int_0^z{\rm d}z^{\prime}
 \sin\psi(z^{\prime})\,
 \E^{{\I}(\Delta_a-\Delta_{\rm pl})\,
 z^{\prime}}c_{\gamma}(z^{\prime})\right|^2
 +\left|\int_0^z{\rm d}z^{\prime}
 \sin\psi(z^{\prime})\,
 \E^{{\I}(\Delta_a-\Delta_{\rm pl})\,z^{\prime}}
 s_{\gamma}(z^{\prime})\right|^2\right)\,.
\end{equation}
\end{widetext}
Here we have assumed $\Delta_{\rm pl}$ to be independent of~$z$.

We next consider a ``patchy'' pattern of domains of equal size $s$
and constant field in each of them. We will show that, when
evaluated after a distance $z\approx N\,s$, with $N\gg 1$, the
conversion probability is roughly the product of the conversion
probability in a single domain times the number of domains. Except
for the replacement $s_\gamma\to c_\gamma$, each one of the two
integrals in Eq.~(\ref{eq:prob2}) can be evaluated as follows, where
$l_0= 2\pi/(\Delta_{\rm pl}-\Delta_a)$,
\begin{eqnarray}
 {\mathcal I}&=&\left|\int_0^z{\rm d}z^{\prime}
 \sin\psi(z^{\prime})\,{\rm e}^{-2\pi {\I}\,z^{\prime}/l_0}
 s_{\gamma}(z^{\prime})\right|^2
 \nonumber\\
 &=& \left|\sum_{k=1}^N\mu_k\int_{z_k}^{z_{k+1}}{\rm d}z^{\prime}\,
 {\rm e}^{-2\pi{\I}\,z^{\prime}/l_0}\right|^2
 \nonumber\\
 &=&\frac{l_0^2}{\pi^2}\,
 \sin^2\left(\frac{\pi\,s}{l_0}\right)
 \left|\sum_{k=1}^N\mu_k\,{\rm e}^{-{\I}\,\pi\,(2z_k+s)/l_0}\right|^2
 \nonumber\\
 &=&\frac{l_0^2}{\pi^2}\sin^2\left(\frac{\pi\,s}{l_0}\right)
 \left(\sum_{k=1}^N\mu_k^2+\sum
 \begin{array}{c}
 {\rm interference}\\
 {\rm terms}
 \end{array}
 \right)\,.\nonumber\\
\end{eqnarray}
Here, $N=z/s\gg 1$ and $\mu_k= |\sin\psi_k|\,s_\gamma(k)$ or
$\mu_k=|\sin\psi_k|\,c_\gamma(k)$ is a random variable in the
interval $[-1,1]$. The random nature of the field directions implies
that the interference term vanishes on average. For the geometrical
factor we have $\langle \mu_k^2\rangle=\langle
\sin^2\psi\,\sin^2\gamma\rangle=1/3$. Then we find
\begin{eqnarray}\label{pertpz}
 P_{\gamma \to a}(z)& \approx &
 \frac{\gag^2 |{\bf B}|^2}{8}\,\frac{l_0^2}{\pi^2}\,
 \sin^2\left(\frac{\pi\,s}{l_0}\right)
 \times 2\times \frac{N}{3}
 \nonumber\\
 &=& N\,\left(\langle\Delta_{a\gamma}\rangle s\right)^2\,
 \frac{\sin^2(|\Delta_{\rm pl}-\Delta_a|\,s/2)}
 {(|\Delta_{\rm pl}-\Delta_a|\,s/2)^2}
 \nonumber\\
 &=& N\,P_0\,,
\end{eqnarray}
having the structure of a probability per single domain $P_0$ times
the number of domains $N$. We stress that Eq.~(\ref{pertpz}) only
holds perturbatively, i.e., $\langle\Delta_{a\gamma}\rangle s\ll 1$
is a necessary condition.

In the limit $|\Delta_{\rm pl}-\Delta_a|\gg
\langle\Delta_{a\gamma}\rangle$, we have in Eq.~(\ref{p1ga}) that
$\Delta_{\rm osc}=|\Delta_{\rm pl}-\Delta_a|$ and Eq.~(\ref{p1ga})
coincides with Eq.~(\ref{pertpz}), provided that $B_T\to\langle|{\bf
B}|\rangle=|{\bf B}|/\sqrt{3}$ because of the projection effect. In
the opposite limit $|\Delta_{\rm pl}-\Delta_a|\ll
\langle\Delta_{a\gamma}\rangle$, Eq.~(\ref{pertpz}) reduces to
\begin{equation}
 P_{\gamma \to a}(z)\approx N(\langle\Delta_{a\gamma}\rangle s)^2,
\end{equation}
again in agreement with the corresponding limit of Eq.~(\ref{p1ga}).

This exercise shows explicitly how the classical rule of ``adding
the probabilities'' instead of amplitudes arises from the randomness
of the polarization and of the field configuration over scales much
larger than $s$. However, since we used first-order perturbation
theory, the validity of these results breaks down when $P_{\gamma
\to a}(z)$ becomes large. This is always the case for $z$ large
enough, since we are not including the back reaction $a\to \gamma$,
that are second order in $\gag$ and that prevent the violation of
unitarity. In the saturation regime, the correct generalization of
Eq.~(\ref{p1ga}) is provided by Eq.~(\ref{totprob}) as discussed in
the text.

\newpage

\end{document}